# Possible Periodic Windowed Behavior in SGR1935+2154 Bursts

Bruce Grossan
UC Berkeley Space Sciences Laboratory, 7 Gauss Way, Berkeley, CA 94720, USA
Energetic Cosmos Laboratory, Nazarbayev University, 53 Kabanbay Batyr Ave., Block C4, office 403, Nur-Sultan 010000, Kazakhstan
Bruce_Grossan@lbl.gov

## Abstract

It has been proposed that repeating fast radio burst (FRB) sources, FRB 20180916B (CHIME/FRB Collaboration, et al., 2020a) and FRB 20121102A (Rajwade, et al., 2020, Cruces, et al., 2021) display periodic windowed behavior (PWB) in the times of FRB detections. In PWB, events occur only within a periodic activity window spanning a fixed fraction of the period, but no events might be detected during many periods. During UT 2020 April 28, two soft gamma burst peaks (detected with Insight-HXMT; Li, et al., 2021) and two FRB peaks (CHIME/FRB Collaboration, et al., 2020b) from soft gamma repeater (SGR) 1935+2154, were detected with arrival delay time between bands consistent with the radio dispersion measure, associating the FRBs and soft gamma bursts. As SGRs are a type of magnetar, a neutron star with an extremely high magnetic field, ~ $10^{14}$ G, these observations establish a link between at least some FRBs and magnetars. The analysis herein shows possible PWB in SGR 1935+2154's soft gamma-ray burst times. For 161 bursts from 2014 through 2020 from IPN (Interplanetary Network) instruments, a resolved minimum in active window fraction vs. period occurs at a 231 ± 9.3 day period, at a 55% active window fraction. The data cover only 6 bursting episodes, however, the periodicity result appears relatively robust: First, the IPN has excellent time coverage and is unlikely to miss bright bursts outside of active windows. Second, for various data subsets (with significantly less bursts), either the best or second best measured period was consistent with 231 days, even in the sub-sample with the richest bursting episode removed. In addition, simulations show that even small numbers of uniform random bursts do not show PWB, and that the results are not due to limited sampling. The appearance of PWB could have alternative explanations, including a non-uniform distribution of burst times; a firm conclusion of PWB would be evident, however, after the observation of more episodes consistent (within a small phase margin) with the given windows. If the periodicity of this SGR's bursts can be verified, and if similar behavior were observed in other repeating FRB sources, it might suggest a linkage between FRB and SGR burst mechanisms or emission conditions.

# 1. Introduction

The year 2020 has been unprecedented in the study of transients, in that the number of cosmological FRB (fast radio burst) detections has exploded, and that observations of a known soft gamma-repeater within our own galaxy, SGR1935+2154, showed two FRBs (or two "components" of FRB 20200428[1], detected near 400 MHz with CHIME, CHIME/FRB Collaboration, et al., 2020b; and near 1.4 GHz with STARE 2; Bochenek, et al., 2020) associated with two soft gamma burst (SGBs) peaks (both peaks detected at 10-250 keV in Insight-HXMT, Li, et al., 2021; detection of at least one peak by Integral IBIS (20-200keV), Mereghetti, et al., 2020; by Agile (18-60 keV), Tavani, et al., 2020, and Konus-Wind (20-500keV), Ridnaia, et al. 2021). The arrival time delay between radio and soft gamma bands is consistent with that predicted by the dispersion measure from the radio observations. SGR sources have been shown to be neutron stars with very high (~ $10^{14}$ G) magnetic fields, magnetars (Kouveliotou, et al., 1998, Kouveliotou, et al., 1999, and e.g. Thompson & Duncan, 1995). So, while there may be important caveats to the precise relation of the SGR1935+2154 event and cosmological FRBs (e.g., Margalit, et. al. 2020), the coincident event proves that FRBs can be made by SGR (magnetar) sources, and that FRBs may be accompanied by SGBs. The latter point raises the emission physics question, “what is the set of physical processes and events that result in both SGBs and FRBs?” One might think first of studying the emission spectrum, or perhaps the time-profile of such bursts, to get at the physics of these events. However, another recent discovery gives a different and perhaps surprising mode of inquiry: periodic windowed behavior (PWB) of repeating FRBs.

In PWB, events are measured only within periodically recurring windows covering some fraction of the period, but many periods may have zero detected bursts. (Note, however, that whether the windows without bursts are 100 % burst-free is not truly known, since the sources have not been observed without interruption for the entire windows at high sensitivity thus far.) PWB has thus far been reported for two repeating FRB sources. FRB 20121102A was found to have P=157±7 days and an active window fraction of 56% by Rajwade, et al., 2020, and by Cruces, et al., 2021 with P=161±5 days and 60% window fraction, with both data sets having no detections in some periods. FRB 20180916B showed P=16.35 ± 0.15 days, and ~25% activity cycle (CHIME/FRB Collaboration, et al., 2020a), again with no bursts reported in several cycles.

Behavior within a periodic window suggests that either some of the causal conditions for the behavior occur periodically, a “production” mechanism, or that conditions for observing the behavior occur periodically, i.e. a “shutter”/alignment mechanism. Most of the magnetars known have a measured pulsation period, believed to be the spin period of the neutron star. These are in the range of 2-12 s (Younes, et al., 2017) and are very far from the windowed periods reported.

[1] Note that FRB 20200428 is referred to in our references as "FRB 200428", due to a change in naming convention.

SGR1935+2154 was discovered in 2014 July (Stamatikos, 2014). Much of the information on this source from observations up to 2016 is given in Younes et al. (2017). It has been detected by several X–ray and gamma-ray instruments, and in radio and in near-IR (Levan, Kouveliotou, & Fruchter, 2018) bands. It is associated with the supernova remnant G57.2+0.8. Additional FRBs have been detected from this source, suggesting a similar behavior to other extragalactic repeating FRB sources (Kirsten, et al., 2020).

# 2. Data

## 2.1. Overview

SGBs are usually reported in the Gamma-ray Coordinates Network (GCN) SGR Archive, and tabulated in the SGR list from the Interplanetary Network (IPN; Hurley, 2007; additional details in the next section). Table 1 offers a summary of IPN bursts for SGR1935+2154. The Table shows that the bursts appear to be in heterogeneous episodes, varying from just a few on 1 day (2014-07), two bursts in 40 days (2015-12 to 2016-02), to a rich episode spread over 105 days (2016-05 to 2016-08). For shutter/align mechanisms (see Introduction), the much shorter or less numerous burst episodes should not be dismissed or given substantially reduced weight. For a more probabilistic view, it may be appropriate to weight bursting activity by episode duration, number of bursts per episode, etc.

*Table 1- Summary of SGR1935+2154 Bursting Activity*

| Episode | Firstdetect | Lastdet | Center | Dur(d) | Summary |
|---|---|---|---|---|---|
| **2014-07** | 2014-07-05 | 07-05 | 07-05 | 0.3 | 4 bursts, single day activity BAT |
| **2015-02 to 04** | 2015-02-22 | 04-12 | 02-27 | 50 | 17 bursts, several instruments |
| **2015-12 to 2016-02** | 2015-12-21 | 02-01 | 01-11 | 43 | 2 bursts on 2 days only, Integral, Fermi, HESS |
| **2016-05to08** | 2016-05-14 | 08-26 | 07-05 | 105 | 81 bursts, several instruments |
| **2019-10to11** | 2019-10-04 | 11-15 | 10-25 | 43 | 20 bursts, several instruments |
| **2020-04to05** | 2020-04-10 | 05-20 | 04-30 | 41 | 37 bursts, 2 Simultaneous FRB, several instruments |

Key: Firstdetect = first incidence of detection in this episode; Lastdet = last detection in this episode; Center = average of burst times; Dur (d) = Duration (time between last and first burst), in days.

## 2.2. Data Selections

In an ideal observation, an instrument with uniform sensitivity to some burst criteria would continuously monitor a given source, and then analyzing all bursts, one could make a definite statement about periodicity or other behaviors within these parameters. In reality, there are many non-idealities to SGR data. A number of instruments operate with widely varying sensitivities, with many inherent periodicities including sensitivity, background, coverage duty cycle, etc. Further, if one instrument detects a bright burst, then other instruments with widely varying sensitivity will frequently schedule observations at some time soon after this, yielding bias toward clusters of bursts after a bright burst. Even this type of coverage is not uniform, however, due to observing constraints. At some level, particularly for faint bursts, the reporting of a burst in acquired data depends on the data reduction, mostly through the burst criteria (but also through background subtraction methods and other, sometimes model-dependent factors).

Many bursts between 2014 and 2016 are reported and analyzed in Younes et al., 2017, and constitute burst episodes beginning 2014 July 5, 2015 February 22, and 2016 May 14, and 2016 June 18. Their displays of bursts vs. date (but not their full analysis) are from bursts detected by IPN instruments, including the Konus instrument (10-770 keV) aboard the *Wind* spacecraft (K-W). K-W is in orbit around the Sun at Lagrange point 1, far from Earth, and so provides a nearly continuous, unobstructed view of the entire sky, an excellent set of properties for time series monitoring. The main downside to these data is that more sensitive instruments exist, and weaker bursts are missed. If shutter/align phenomena are of primary interest, it may be more important to include all bursts than to have a uniform sensitivity. One cluster of bursts from K-W, on 2019 Nov 4 and 5 (Ridnaia, et al., 2019), was not reported by more sensitive wide-field instruments, including Fermi-GBM. This demonstrates the almost complete time coverage of this instrument, even though the sensitivity is limited. It also demonstrates the problem of sensitivity vs. coverage; if other instruments covered the event, they might have found weaker bursts covering a wider time window.

All confirmed bursts from the source detected by any spacecraft in the IPN (including K-W, Fermi GBM, and Swift-BAT) up to 2020 Dec. 20 were obtained from the IPN website (Hurley, 2007) SGR burst list for this analysis. The following data selections were analyzed:

**All IPN bursts through 2020 Dec 13** (IPN20) – All IPN bursts as given above. Total=161 bursts.

**K-W detected bursts, plus some likely additional** (KW+) – A search was made of the GCN circulars archive for SGR1935+2154 bursts. The correspondence between GCN reported K-W bursts, and IPN reported K-W bursts is one-to-one except on April 27. GCN 27667 (Ridnaia, et al., 2020a) gives "tens of bursts" detected by K-W on this date; the IPN gives only 4 detected by K-W, but a total of 18 by all instruments. All 18 bursts are therefore counted for this selection. Total=39 bursts.

**Unrestricted Equal Weight Episodes** (UEW) – The median episode duration, 43 days was used, and one burst per day was assigned to 43 days around the average center time of all 6 episodes. (This is an artificially constructed data set intended to explore the concept of a periodic shutter mechanism modulating events.) Total=258 bursts.

We offer two additional samples in order to test the sensitivity of our result to removing a single large burst episode. **IPNA** is the IPN20 selection with the mid-2016 episode, our richest episode, with the largest number of days with bursts, removed. Total=80 bursts. **IPNB** is the IPN20 selection with the 2020 episode removed. Total=124 bursts.

The bursts in our selections are listed in the Appendix.

# 3. Methods: Analysis of PWB

Periodic behavior, especially in pulsar studies, is typically studied using Fourier analysis, or Lomb-Scargle periodograms, particularly used for uneven sampling (e.g., VanderPlas, 2018). Folding data into finite phase bins at a trial period (e.g. Staelin, 1969) has been used to evaluate periodic behavior in different contexts, and various methods are used to determine the best period, such as maximizing the chi-squared, (that is, the sum of the data minus the global average among the phase bins, squared, divided by sigma squared) and the minimum string length method (e.g., Clarke, 2002). These methods were developed to identify truly periodic signals, not to identify activity windows, however.

As we are looking for PWB, instead, the data were folded at a range of periods, with the best period chosen to be the one that minimized the size of the window that contained 100% of the bursts, the W100 statistic, expressed as a fraction of the period. It is important to note that this statistic is calculated without regard to the profile peak, i.e. it is not symmetric about the folded profile peak. This is essentially the same method used in Rajwade et al. (2020). The data were folded at trial periods in single day steps and 10 phase bins (the chi-squared statistic is not used for any selection here, but it is calculated for this binning). Both the minimum W100 widths and the presence of any obvious "dip structure" in the W100 vs. period plot were noted. Three of the four burst episodes occur within about 800 days, therefore, period durations were searched between about 100 and 400 days. This procedure yielded clear, resolved W100 minima with our data. FFT analysis and folding with chi-squared minimization did not yield as clearly preferred periods; most importantly, these latter analyses are not directly related to the window width we are looking for (i.e. W100).

# 4. Results

## 4.1. Event Analysis

The results of period folding and W100 measurements are summarized in Table 2. Figures 1-3 show the W100 vs. period plots for some of the data selections. Figure 1 shows, for the IPN20 data selection, the largest data set, that a minimum of W100, 55%, occurs near the center of a resolved, smooth dip, at a period of 231 ± 9.3 days; this period therefore best describes the PWB in the SGBs of this source. (The uncertainty reported is the half-width of the W100 minimum, following Rajwade, et al., 2020. The dip is somewhat asymmetric but the uncertainty generously covers the difference between the dip center and the minimum). Figure 4 shows the data and derived periodic windows.

For the IPN A selection, with the episode with the most bursts removed, the analysis gave two almost equally deep W100 minima. The lowest W100 minimum was at 268 ± 12 days, W100 minimum =45%. However, the 231 day period is the center of the second lowest dip, with W100 minimum = 48%, and the depth of this dip below the local "continuum" is within 2% of that at 268 days. The IPN B selection with the 2020 episode removed has the preferred 231 day period (231 ± 14 days) for the W100 minimum. In the K-W+ data selection, dips are present at 203 ± 7.5 days, with W100 minimum=32%, and 238 ± 6.9 days, with W100 minimum = 34%, the latter consistent with the preferred 231 day period. The depth of the dips below the local "continuum" are different by ~ 10%.

*Table 2  Samples, Trial Periods, Width of the Window Containing 100% of Bursts, W100, and Chi-squared.*

| Sample | Period(d) | W100 | Chi-sq |
|---|---|---|---|
| ***IPN20** | 231 | 0.554 | 219.8 |
| IPN20 | 232 | 0.556 | 220.0 |
| IPN20 | 233 | 0.558 | 187.5 |
| IPN20(c) | 234 | 0.560 | 248.4 |
| IPN20 | 235 | 0.562 | 243.9 |
| *IPN20 | 277 | 0.628 | 157.8 |
| (c)IPN20 | 280 | 0.632 | 231.1 |
| *(c)IPN20 | 393 | 0.642 | 158.7 |
| ***IPNA** | 268 | 0.453 | 114.9 |
| IPNA | 269 | 0.455 | 136.0 |
| IPNA | 270 | 0.457 | 139.4 |
| IPNA | 271 | 0.459 | 137.4 |
| (c)IPNA | 272 | 0.461 | 103.9 |
| *IPNA | 233 | 0.475 | 94.8 |
| (c)IPNA | 231 | 0.493 | 100.8 |
| *(c)IPNA | 153 | 0.540 | 98.4 |
| ***IPNB** | 231 | 0.554 | 147.5 |
| IPNB | 232 | 0.556 | 145.9 |
| IPNB | 233 | 0.558 | 173.6 |
| IPNB | 234 | 0.560 | 174.4 |
| IPNB | 235 | 0.562 | 160.2 |
| (c)IPNB | 239 | 0.569 | 193.5 |
| *IPNB | 277 | 0.628 | 145.1 |
| (c)IPNB | 282 | 0.635 | 145.3 |
| *(c)IPNB | 392 | 0.639 | 162.6 |
| ***(c)KW+** | 203 | 0.315 | 23.3 |
| KW+ | 202 | 0.316 | 20.0 |
| KW+ | 201 | 0.318 | 19.1 |
| KW+ | 204 | 0.338 | 27.6 |
| KW+ | 200 | 0.339 | 20.9 |
| *(c)KW+ | 238 | 0.340 | 18.6 |
| *(c)KW+ | 155 | 0.412 | 19.0 |
| ***UEW** | 236 | 0.503 | 269.1 |
| (c)UEW | 235 | 0.514 | 266.0 |
| UEW | 237 | 0.524 | 265.1 |
| UEW | 234 | 0.525 | 252.1 |
| UEW | 233 | 0.536 | 238.0 |
| *UEW | 177 | 0.545 | 243.0 |
| (c)UEW | 175 | 0.562 | 240.6 |
| *UEW | 278 | 0.567 | 229.1 |
| (c)UEW | 280 | 0.582 | 237.2 |

The table above lists the lowest 5 W100 values for each data selection, plus the approximate center (c) and minimum (*) values of the next two lowest resolved dips in W100. Note the dip center and minimum can be the same.

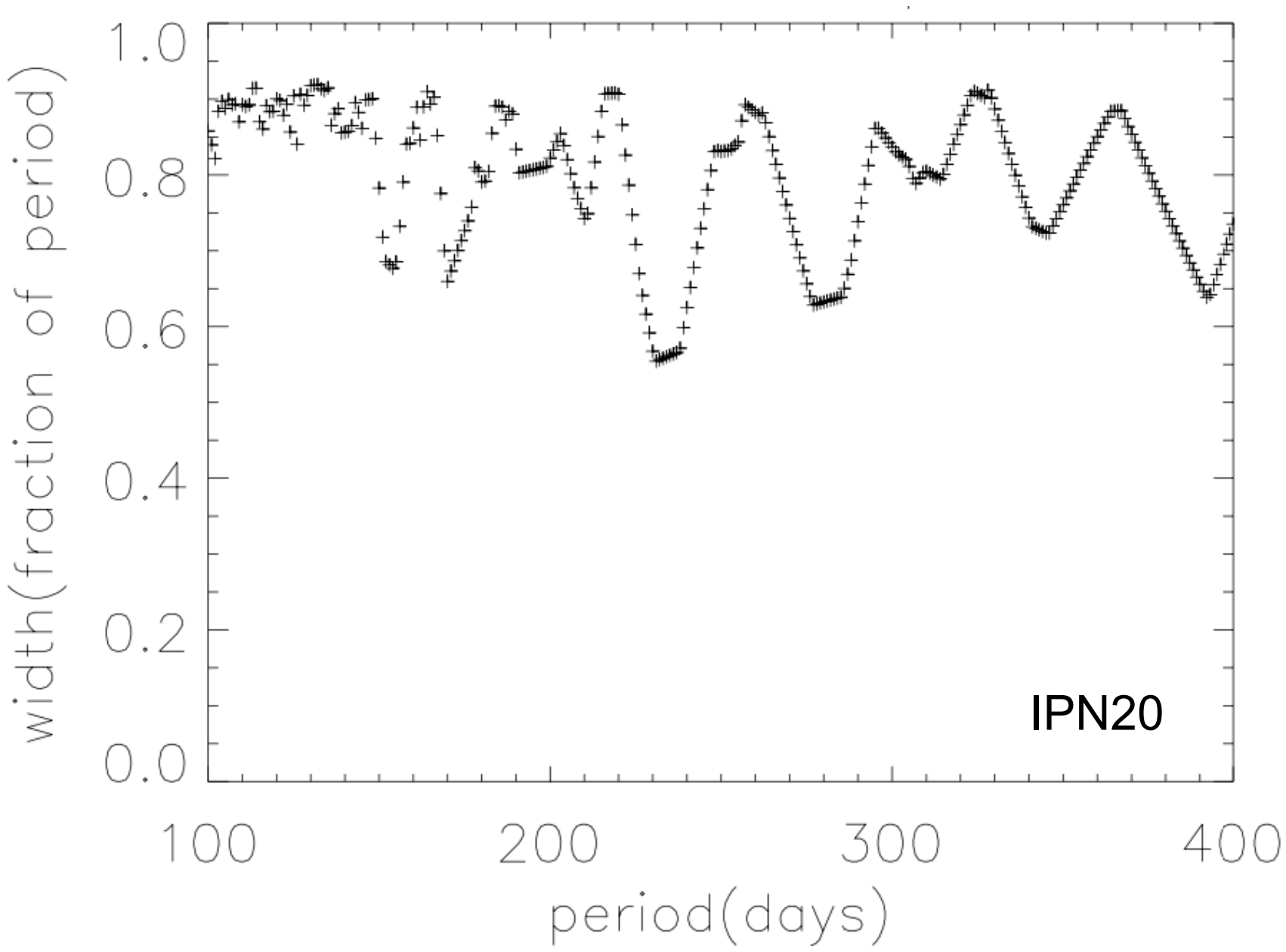

Figure 1 Periodic window fractional Width (W100) vs. period for the IPN20 selection. The minimum W100, or the width of a periodic window containing 100% of the bursts, gives the preferred 231 ± 9.3 day period from this, the largest data set.

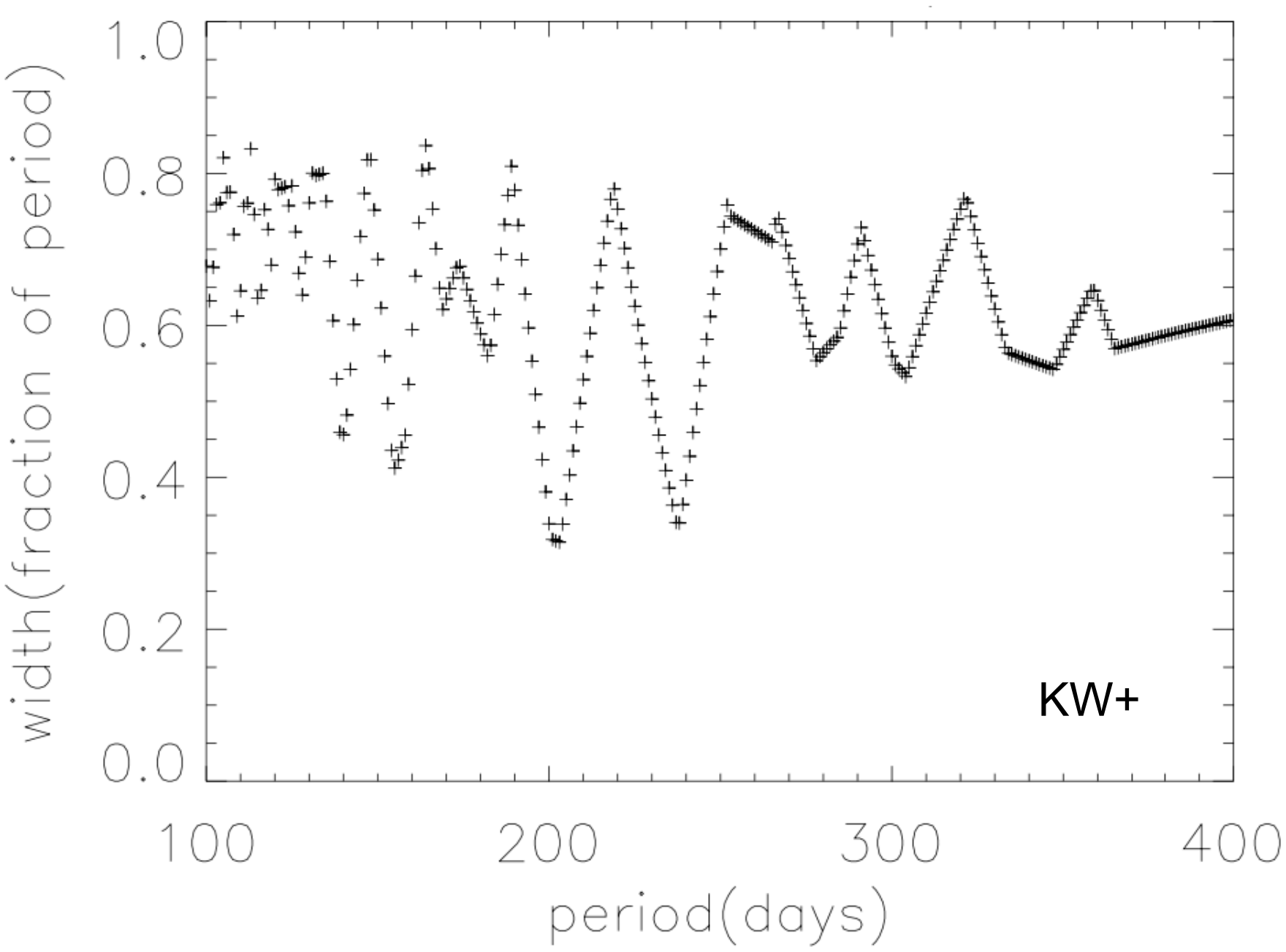

Figure 2 W100 vs. period for the KW+ selection. The marginally deepest minimum is at 203 ± 7.5 days, but another, almost equally deep, is present at 238 ± 6.9 days, near the nominal period. This data sample is much smaller than the previous.

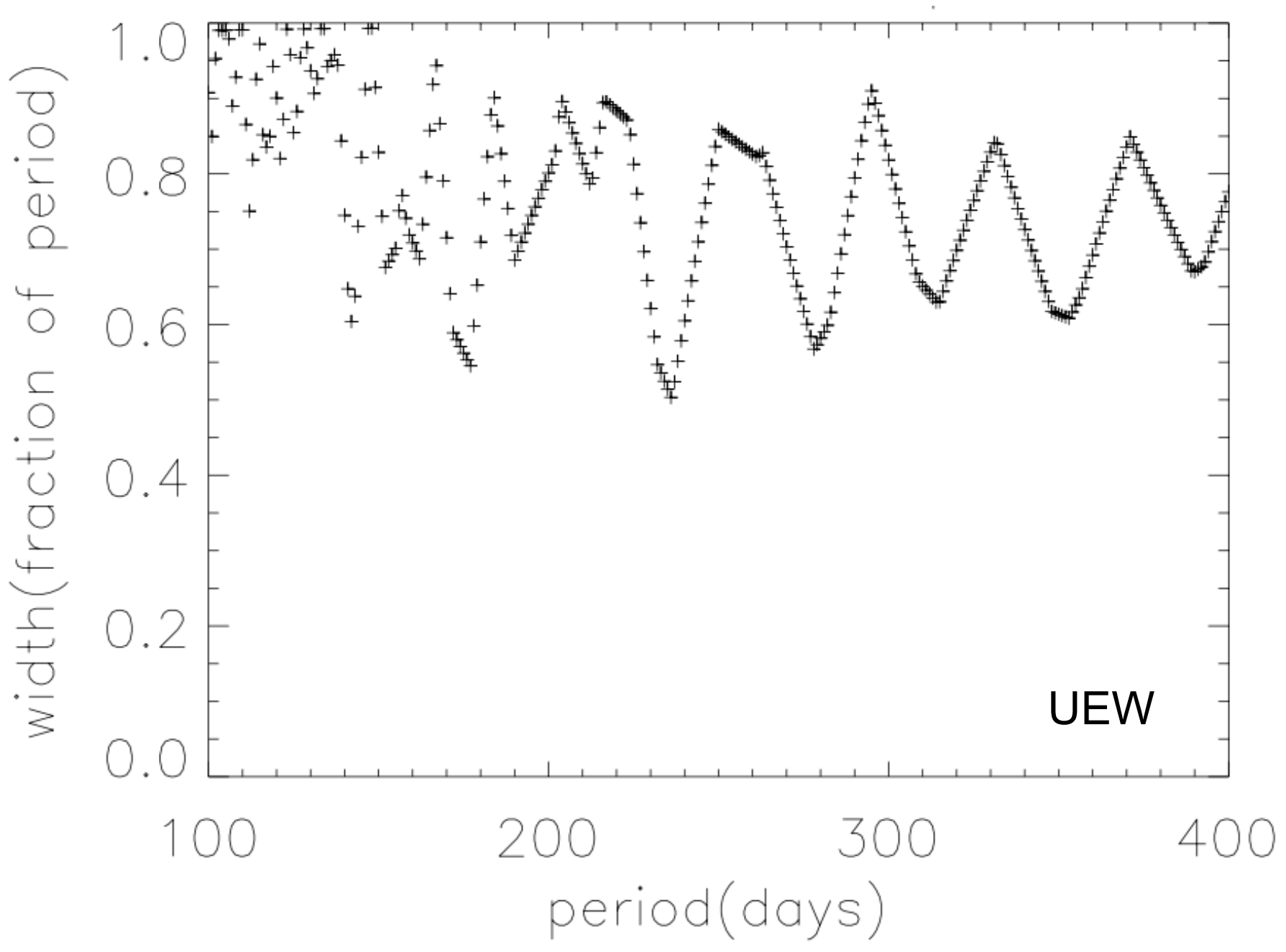

Figure 3  W100 vs. period for the UEW selection. There is a prominent dip at 236 ± 7.3 days, also near the W100 minimum value for the IPN20 selection.

The folded data give a "pulse profile", shown in Figure 5 for the IPN20 data selection. This profile is not like a pulsar pulse profile or other periodic behavior profile, as the behavior is not known to be strictly periodic; it should not suggest there is a clear burst probability vs. phase function. There are simply not enough data within each window (just two points in some episodes) to draw a significant conclusion about structure within the window. The profile could change with future data, even if such data are consistent with the activity windows we find here. The profile does show a peak, however. This peak is at the same phase in all analyses at the 231 day period (and in the minimum W100 period for IPNB), except in the UEW sample, which is expected to remove profile structure. Note that the UEW selection still yields almost the same resulting W100 minimum period (236 ± 7.3) as the IPN20 selection (Figure 6).

# 5. Comparison to random and Weibull distributed data

## 5.1. Random Data Simulations

The analysis of random data with similar general characteristics to the actual data helps assess the veracity of the result. The random data were reduced with the same software as the real data. Many realizations of simulated data sets were made and analyzed, with events distributed at random (uniformly) over 2400 days, about the same time span from the first to the last burst. Analysis of these simulated event streams shows that the larger the number of burst events, the less likely small W100 values will be produced, i.e. the less likely significant PWB will be found. The probability of W100 ≤55% is ~4 $\times 10^{-5}$ for 32 events. The probability of a false periodicity and similar W100 for 161 events, the same number as in the IPN20 sample, is vanishingly small, and could not be determined with only the ~$10^5$ trials used here.

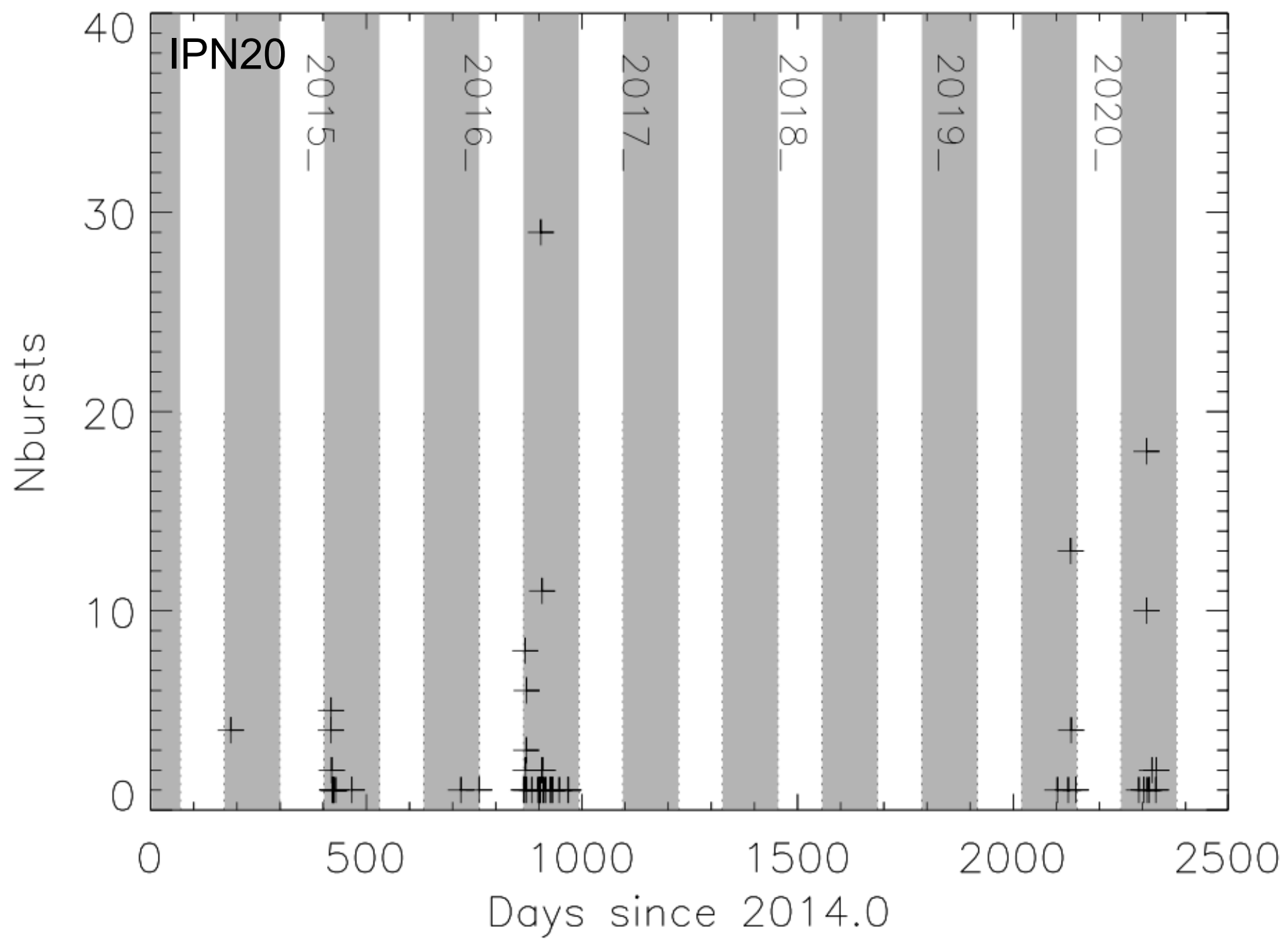

Figure 4 IPN20 selection bursts and windows. The figure shows the IPN20 data, binned into single days, and shaded active windows with a 231 day period, corresponding to the minimum W100 of 55%. The data selections are described in Section 2.2.

### 5.2. Periodically Windowed Random Data

Assume the hypothesis that the true phenomenon is random events that either only occur, or are only observable, during some of the periods, and only within periodic windows. In this case, analysis of simulated data can demonstrate the limitations of measurement of PWB periods and W100. In particular, one might anticipate that with small numbers of events, it would be difficult to determine the period or window fraction. Random (uniformly) distributed events were therefore generated with the same characteristics as the real data, i.e. near the claimed 231 day period and 55% window fraction, and activity in only 6 of 11 periods. Sets of such randomly generated data, with different numbers of total data points, were therefore generated to assess the quality of the results of our analysis.

The period determination turned out to be very good as long as more than about 43 events were in the data set (fractional variation of ~ percent; Figure 7). With only a few times $10^3$ realizations and 43 events, we could not find any instance of the W100 method more than a few percent off the true period. The W100 value itself, however, showed significant variation (always underestimating) until about 100 points, where the errors were < 5% (Figure 8a). This can be understood by imagining starting with a data set with good, uniform coverage, then removing points at random one-by-one. Eventually, the extrema points of each window would be removed, and the W100 would decrease. However, the period would still be present in the data. (The folded profile, a square wave with large numbers of points, would lose non-zero values from more phase bins as the number of points was reduced.)

Random windowed data were also reduced by chi-square maximization. Period values with very large (~50%) errors were present for 86 points (compare to a clear convergence and less than 5% errors for 43 events with W100 minimization). For window fractions, the W100 method gave errors < 10% at 129 events; chi-squared method errors did not continue to decrease after 43 events, and errors never became small. This demonstrates that the chi-squared technique does not work well for PWB, though regularly used for true periodic behavior.

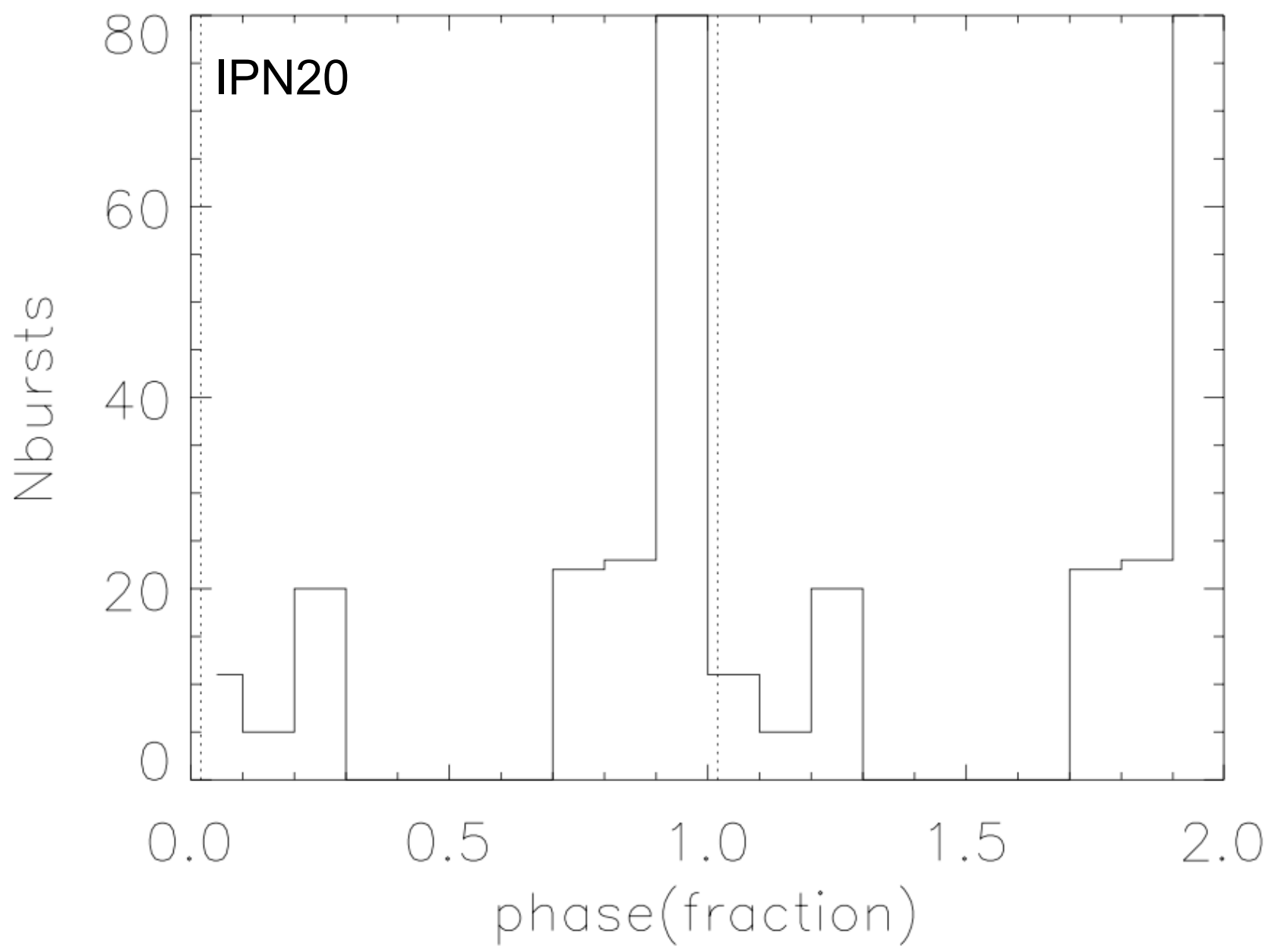

Figure 5 Folded profile at 231 day period, IPN20 data selection. A dotted vertical line indicates the center (midpoint of boundaries) of the W100 window. (Note the profile is repeated twice, following astronomical convention. Also note that zero phase is defined from 2014.0, not from the beginning of the active window. In the IPN20 sample, the active phase is from 0.741 to 1.296.).

### 5.3. Weibull distributed data and clustering

The Weibull distribution was used to investigate clustering of FRB bursts within episodes (Cruces, et al., 2021; Oppermann, Yu, & Pen, 2018). Could such "clustered" behavior explain PWB shown for the data presented here? Following previous works, the cumulative density function (CDF) of time intervals between bursts ($t_i$-$t_{i-1}$) was formed for the IPN20 data set, and compared to that for a Weibull distribution. Note that the CDF is independent of any binning choice; the data define the location of each value in the distribution. The data CDF flattens at ~ 20 –150 days, showing that most adjacent points are within the same active window, and so separated by a small fraction of its duration (55% of 231 days =127 days). A fit of this CDF to that of a Weibull distribution was made, with $CDF(x)=1-e^{-(x/\lambda)^k}$ , $x > 0$ ; 0 otherwise; $x$ = time interval between bursts; λ and k the "scale" and "shape" parameters, respectively. In order to concentrate on longer time scales relevant to the PWB found here, a fit was made of intervals only > 2 days. The best fit, shown in Figure 9, is poor, and the uncertainties on the parameters are extremely large. It was possible to make a creditable fit to small parts of the CDF separately, but no fit would reasonably follow all of these regions, as shown in the Figure. The Weibull distribution, like other smooth clustering distributions, cannot be readily "tuned" to the specific time scales present in the data. Weibull distributed random is also extremely unlikely to produce PWB; in $10^4$ realizations of simulated data at the best fit parameters, for the same number of bursts as in the IPN20 selection, no W100 < 94% was found (and in thousands of additional trials for a surrounding range of parameter values, no W100 < 92% was found).

# 6. Discussion

## 6.1. Robustness of Results and Alternative Interpretations

The data cover only six bursting episodes; one consists of only two burst detections, and another covers only one day. This might, at first glance, throw into question the robustness of these results. However, several points give us confidence that little will change with re-analysis or additional data in the epochs we cover: First, de-selecting significant amounts of the data often had little effect on the results. Removal of even the richest burst episode, the IPN A selection, yielded a consistent result, in that two dips in W100 were present, of essentially the same depth, one at the preferred period. Removal of the rich 2020 episode, the IPN B selection, yielded the preferred period. Next, it is unlikely that, down to the sensitivity of K-W, any bursts have been missed between the given windows. Where required for localization, the IPN also includes the ensemble coverage of Fermi GBM (all-sky, but with Earth occultation; NASA, 2011), BAT (~2.2 Sr; Barthelmy, 2004), and others. The time that K-W plus other instruments would not yield a required localization is small. Finally, the random simulations show that any W100 ≤ our nominal is very unlikely unless ~32 events or fewer are used. This means that the skeptic may eliminate any of our data for any type of test or different analysis, but unless the data are reduced by a factor of ~5, random data are extremely unlikely to produce a W100 as small. Finally, if we assume the hypothesis of windowed periodic behavior is correct, and events are distributed uniformly within the active windows, the number of data points available suggest errors on our W100 and period values of less than a few percent. Taken together, these points suggest a robust result with the data available.

An unbiased examination of these results must allow that PWB is not a unique interpretation of the data; finite statistics and limited knowledge of the physical system under observation allows other periodic functions (e.g. a periodic burst probability function with high and low probability phases of the cycle near to our active and inactive windows) and even non-periodic functions that could mimic our data. In the latter category, one might imagine a clustered random burst distribution; such a distribution that could produce the many bursts observed within an episode, and the right timescale between episodes could reproduce the apparent PWB. This would require significant "tuning" to produce the correct episode and in-between episode time scales, however. In particular, as the K-W instrument gives the IPN nearly all-time and all-sky coverage, essentially no bright bursts could be missed; the suppression of bursts on other timescales (tuning) must be extreme. As was pointed out in Section 5.3, simulated Weibull clustered data with "best fit" parameters could not produce PWB similar to that measured by the analysis herein. However, this analysis only formally eliminated Weibull distributed data; it did not rule out other types of distributions. Additionally, if one considers only the observation of 6 episodes, not considering the actual bursts (and ignoring the rich information therein), considerable room is left for other models, even random occurrence. The random probability of 6 episodes in 10 periods occurring only within 6 of the 55% active windows, (a rough approximation to our data) is about $7.6 \times 10^{-3}$, a not inconsiderable probability; this does not include the occurrence of more than 6 episodes within 6 windows, which could also reproduce the observed pattern of episodes, adding additional, but small, probability. So, other scenarios could produce similar behavior of episodes, possibly with non-negligible probability. This PWB analysis, however, is motivated by the observation of PWB in other FRB sources (CHIME/FRB Collaboration, 2020a; Cruces, et al., 2021; Rajwade, et al., 2020), albeit in radio bursts. The approach herein should therefore be useful and relevant, even if additional confirming data are required to eliminate alternative interpretations.

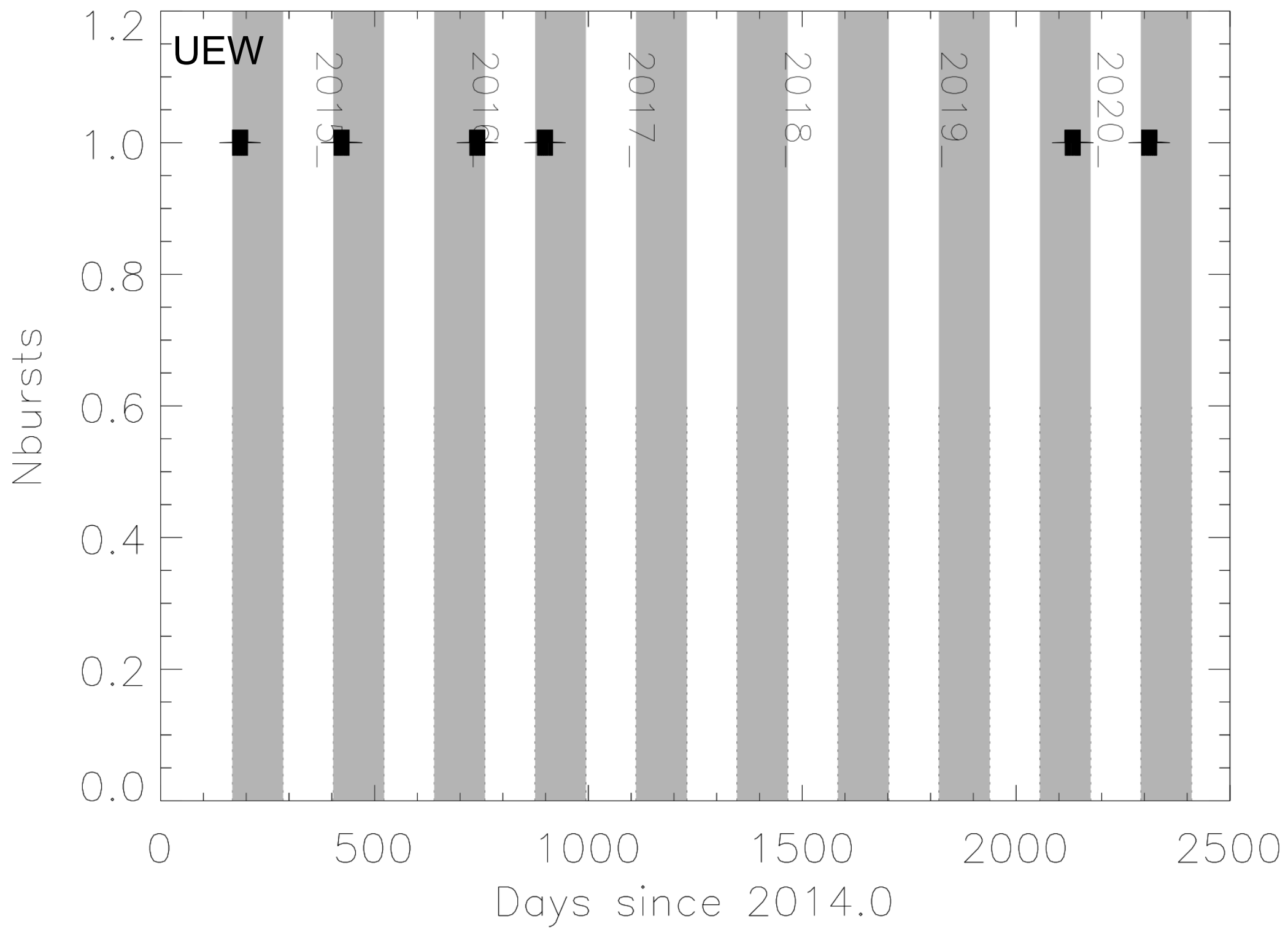

Figure 6 Bursts and windows for the UEW selection. The figure shows the UEW data selection, binned into single days, and shaded active windows with the 236 day period, corresponding to the minimum W100 of 50%. This period is very close to the result for the IPN20 period, 231 days.

## 6.2. The Origin of Windowed Periodicity

Windowed periodicity, as proposed here and in repeating FRBs, is not as yet associated with a physical mechanism. At first glance, one might wonder if the influence of an orbital companion might produce the periodic behavior. However, no magnetar companions are reported in the literature. Magnetars are almost by definition isolated neutron stars (NS) powered by magnetic energy (e.g. Mereghetti, 2008) and have distinct properties from X-ray binaries. No accretion activity is proposed for SGRs, unlike X-ray binaries. There are FRB models that have a companion and no accretion (Zhang & Gao, 2020, and references therein), including the Binary Comb Model (Ioka & Zhang, 2020). This model produces FRBs in the interaction of a B star wind and a NS magnetosphere. In using this model to explain the 16.35 day periodicity in FRB 20180916B (Zhang & Gao, 2020), it was found that binaries with periods from 1-100 days would form, but "after 100 days the birth rate significantly decays". As the periods for SGR1935+2154 SGBs, and for FRB 20121102A FRBs are significantly larger than 100 days, either the model needs modification or it does not explain these objects. It must be noted, however, that free precession of NSs could be the origin of some kind of periodic behavior, without the presence of a companion (e.g., Pines, 1974; Akgün, Link, & Wasserman, 2006).

Models explicitly addressing simultaneous FRB/SGR events may invoke pair production involved in the SGBs (generally believed to originate in disturbances in the neutron star crust) and a magnetospheric interaction to produce the FRB relatively nearby the surface (Kumar & Bošnjak 2020; Kumar et. al. 2017), or e.g. a jet and shock model, that emits the radio burst relatively farther from the NS surface (Margalit, et. al. 2020). These models do not explicitly consider periodicity or a companion, however. A periodic phenomenon might modulate the orientation or structure of the magnetic field affecting magnetospheric phenomenon or jet containment. A mechanism for affecting the magnetic field might be some motion of an accretion disk, as ionized matter can drag field lines. Early discussions of SGR activity proposed some kind of accretion disk, possibly a remnant (e.g. Mereghetti, 2008). Though the Keplerian periods of objects near a NS are far shorter than 231 days, it is conceivable that this periodicity could be some kind of “beat frequency” with another period, e.g. an orbital period and some period of disk motion that are close, but not matching. There is, however, no evidence of any such disk, and early relic accretion disk models for SGRs are no longer favored (i.e. not mentioned in the recent references herein).

Another origin of periodicity is in regulation of the visibility of emission, a “shutter” or alignment mechanism, rather than in the emission itself. How a periodic phenomenon might produce conditions favorable for viewing soft gamma and FRBs, or periodically remove an obscuration to favor observation, is not obvious. Note that to obscure gamma-rays a substantial column of gas or electrons would be required.

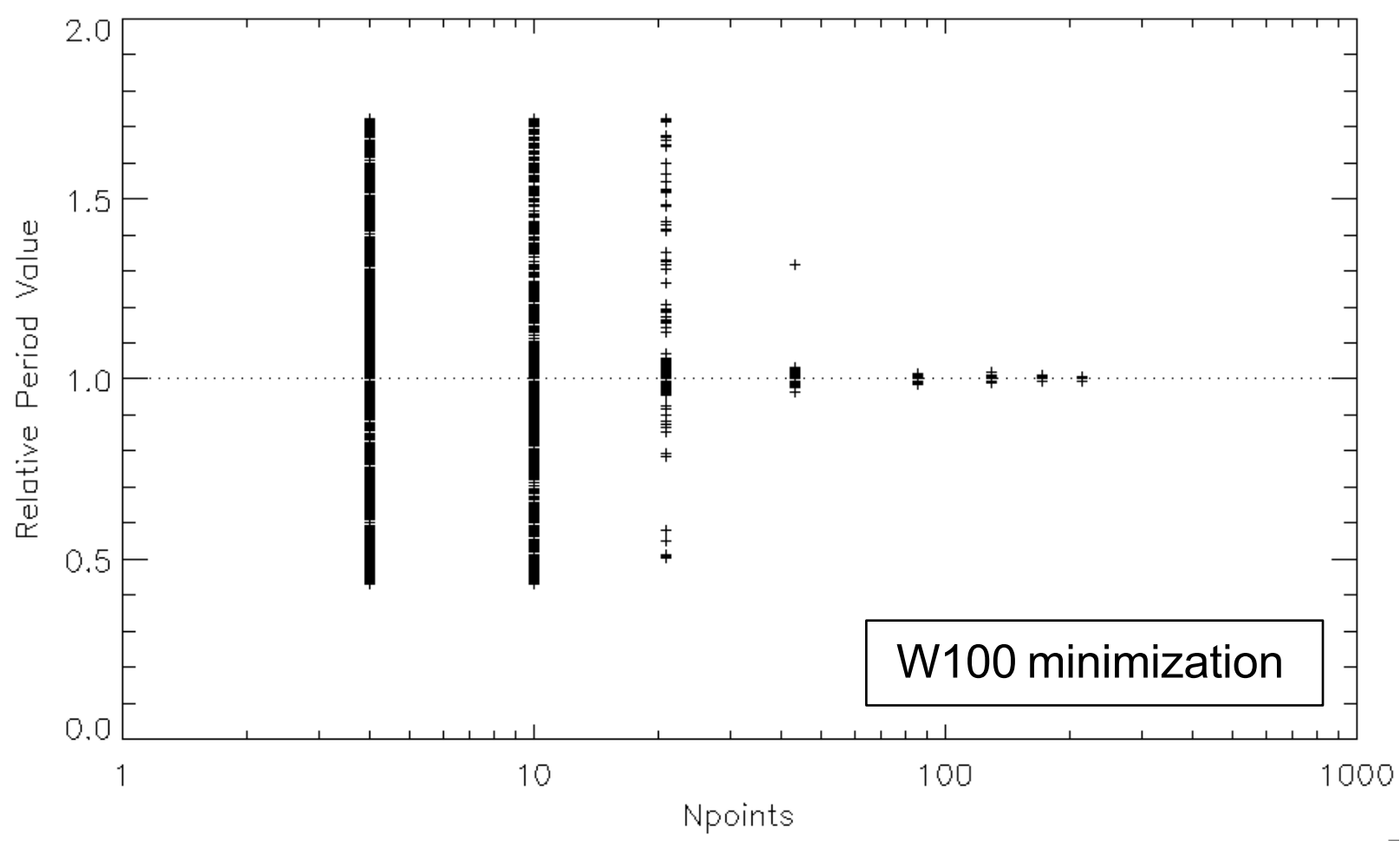

Figure 7a. Measured period values from random events in periodic windows. At 43 events or more, the correct period is determined with <1% error; at 86 events no outliers are found. (The apparent small number of points at large $N_{points}$ is due to integer period values; 2500 realizations were made for each $N_{points}$ value.)

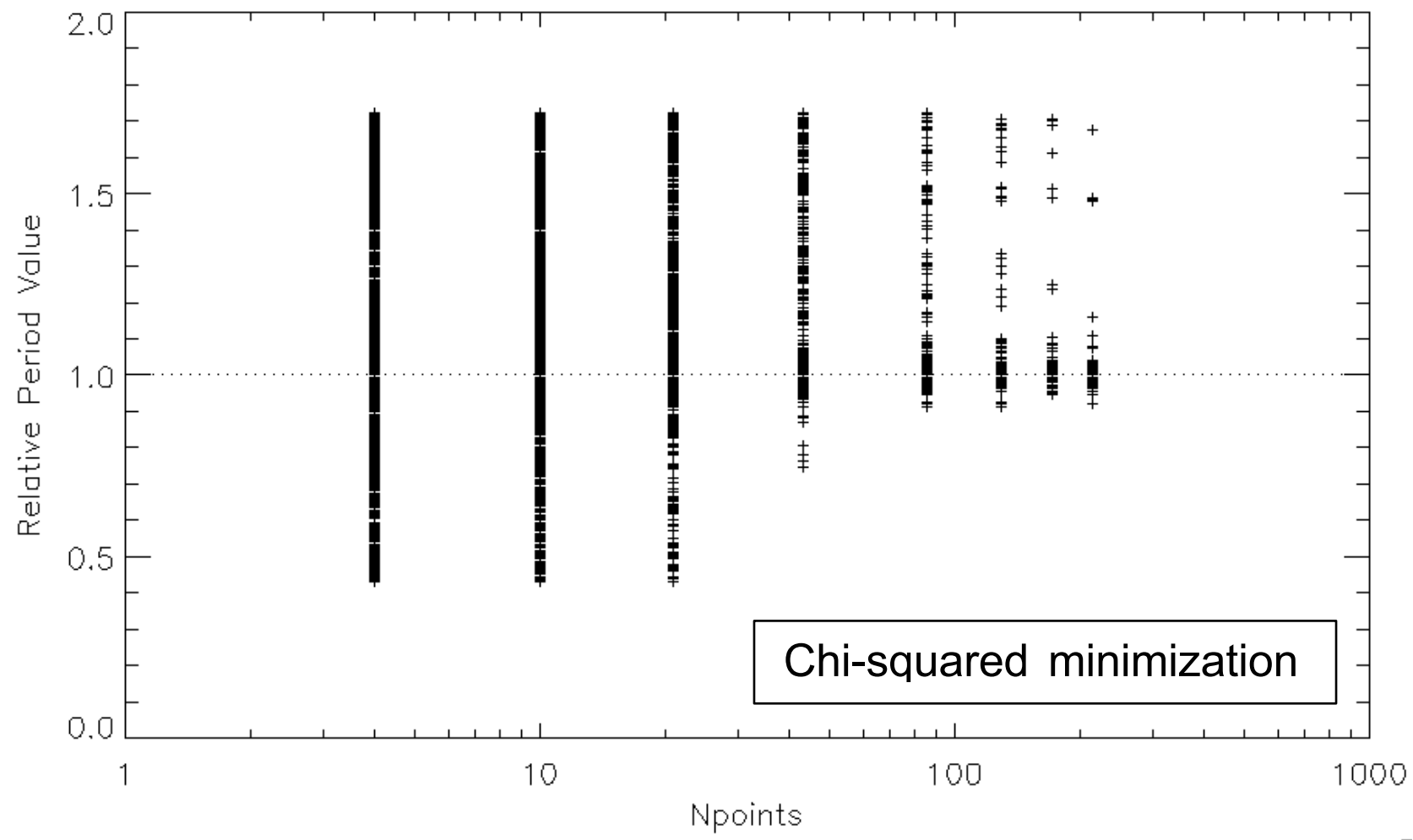

Figure 7b. Measured period values from random events in periodic windows by chi-squared maximization method. The variance in the period is much larger than for the W100 method, and even for 215 data points, there are still alarming outlier results. The same number, 2500 realizations were made.

## 7. Conclusion

A folding analysis was used to show possible PWB in the broadly defined SGBs of SGR1935+2154. It is extremely unlikely that this result would come from insufficient or irregular sampling of uniform random data. It was also shown that for simple assumed PWB composed of random events with the given period and window fraction, the number of bursts makes it unlikely that the period could be far off the correct result. It was also shown that a conventional chi-squared analysis is poor at identifying PWB. As pure speculation, given the more limited tools for analysis of PWB, compared to those for true periodicity (especially pulsar analysis), it could be that PWB is not uncommon in various types of astrophysical data, but rarely recognized thus far.

Alternative interpretations to PWB must be considered, however, including that of "clustered" behavior, given that the data are distributed in only 6 "episodes" or active windows. As shown above, such behavior is not consistent with commonly examined smooth clustering distributions (e.g. the Weibull distribution). However, a distribution that preferred the episode length measured herein, and suppressed time scales in between that and the measured period, cannot be eliminated with the current data. Future data (unless the source ceases activity) will verify or refute the PWB description: As more episodes are measured within predicted windows, other interpretations (such as clustered bursts) would become untenable; conversely, if future analysis of additional bursts (with similar burst properties) yields much larger W100 minima, the PWB interpretation would become untenable. The next 5 windows predicted by our current analysis at the preferred 231 day period are given in Table 3. Due to the apparent randomness of burst times, it is likely that W100 would increase somewhat with more bursts, and the period might change within the given uncertainty. However, if any significant number of similar bursts were detected near the middle of an inactive window, this would invalidate 231 day PWB. If, with more data, only small numbers of bursts near the middle of the proposed inactive window significantly increased the W100 minimum value, but other measures still showed periodic behavior, a model of a periodically modulated burst probability, rather than strict windows, might be more appropriate than the PWB proposed here. (Note added post-submission: Bursts detected after the IPN SGR list here, in 2021 Jan. and Feb., are consistent with the predictions in Table 3; Ricciarini, et al., 2021, and Kawamuro, et al., 2021.)

*Table 3 Predicted active window start and end dates.*

| Window Start | Window End |
|---|---|
| 2020 Oct 17 | 2021 Feb 22 |
| 2021 Jun 5 | 2021 Oct 11 |
| 2022 Jan 22 | 2022 May 30 |
| 2022 Sep 10 | 2023 Jan 16 |
| 2023 Apr 29 | 2023 Sep 4 |

Though the analysis result here appears robust with the data available, no identification with any known mechanism for the PWB was found. This work in no way addressed burst characteristics such as duration or fluence, nor spectral properties. These are likely physically important, e.g., in GCN 27669 reporting the detection of an SGB coincident with an FRB (Ridnaia, et al., 2020b), this observation was offered: “The burst temporal structure and hardness differ from a typical SGR burst … suggesting a possibly different emission mechanism.” Perhaps only a tiny subset of the SGBs studied here are associated with the same physical mechanism of those coincident with FRBs. Though this source produces both FRBs and SGBs, and though some FRB sources show FRB PWB, the proposed SGR PWB here is not a sufficient demonstration of a direct or causal link between the two physical mechanisms. However, if one allows that PWB is a common behavior in the one known Galactic FRB source plus several extragalactic repeating FRB sources, then it would seem reasonable (though not logically proven) that PWB is likely a general characteristic of FRB-causing systems. This characteristic might then be a required feature of future models of magnetars, restricting the variety of possible models, and steering us toward a proper identification of the physical mechanism behind FRBs. The author urges additional observations and analyses covering significantly more putative periods of this SGR, in order to clarify the nature of this possible periodicity, as well as increased monitoring of other SGRs, to potentially progress toward a physical understanding of SGR sources and their connection to FRBs.

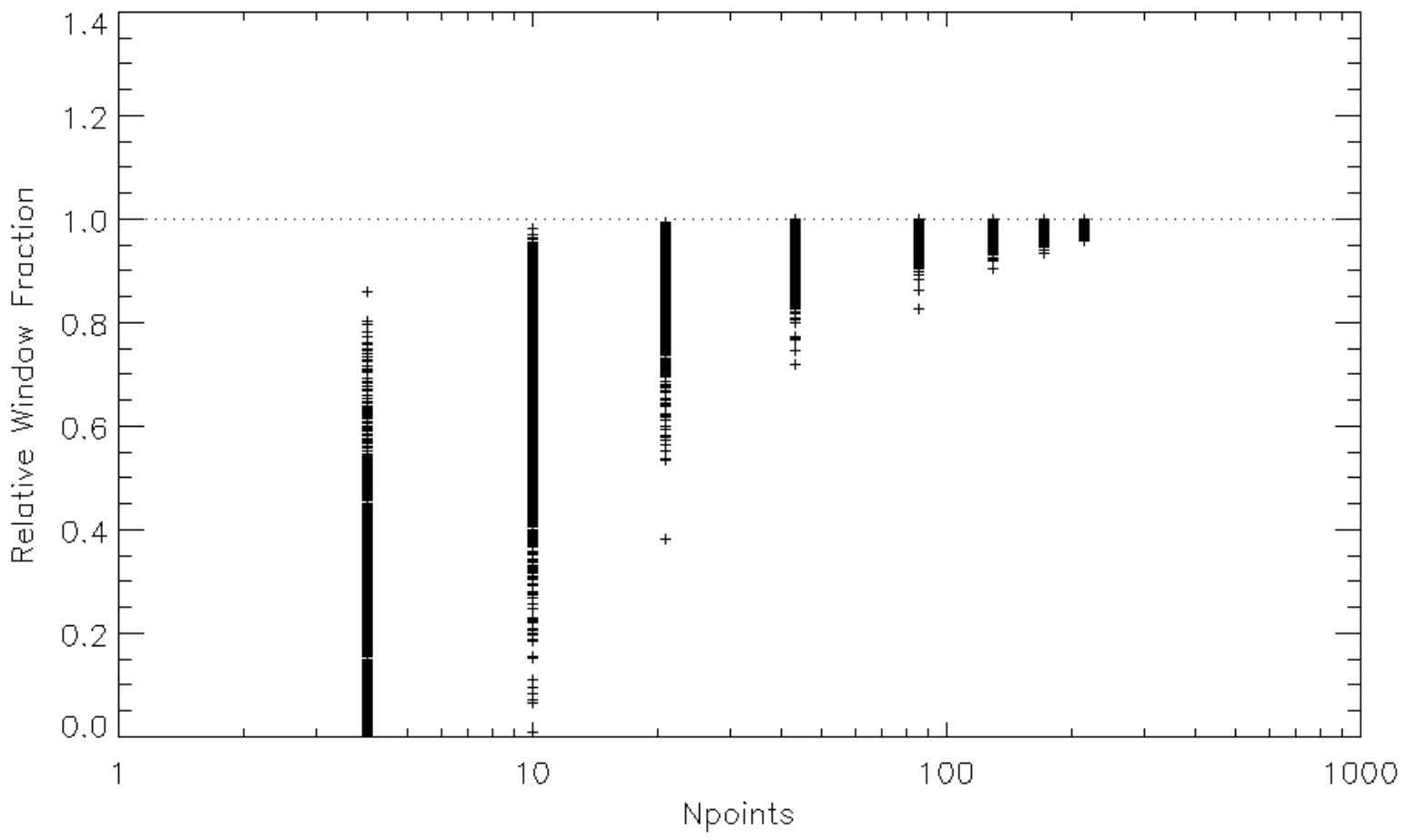

Figure 8a Measured window fraction from random events in periodic windows, by W100 minimization. The standard deviation in the minimum window fraction is less than 4% with 43 points (but biased low). Each $N_{points}$ is 2500 realizations.

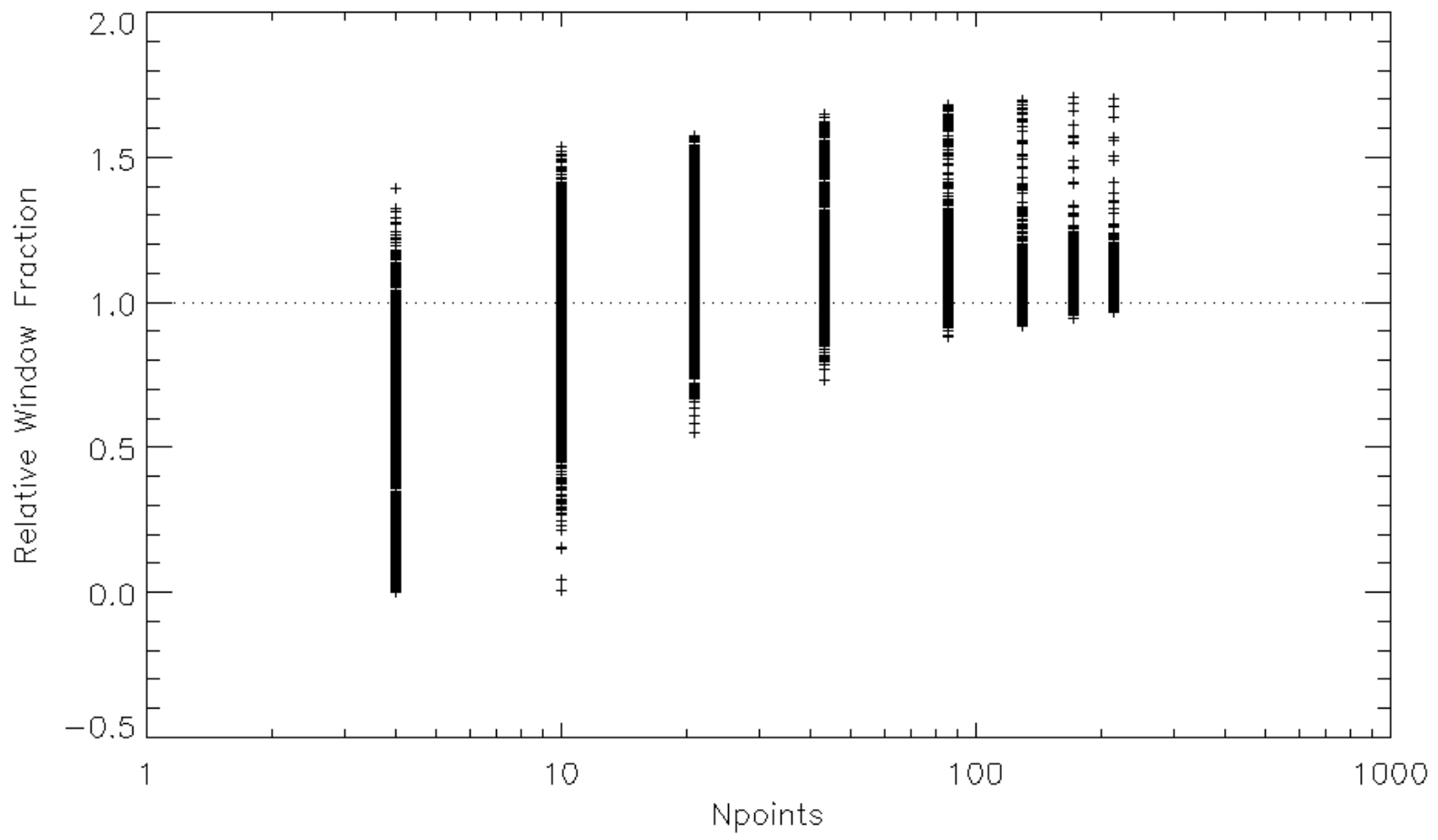

Figure 8b. Measured window fraction from random data, by chi-square maximization method. Even above 200 points, the errors are large and outliers remain. Also 2500 realizations for each $N_{points}$.

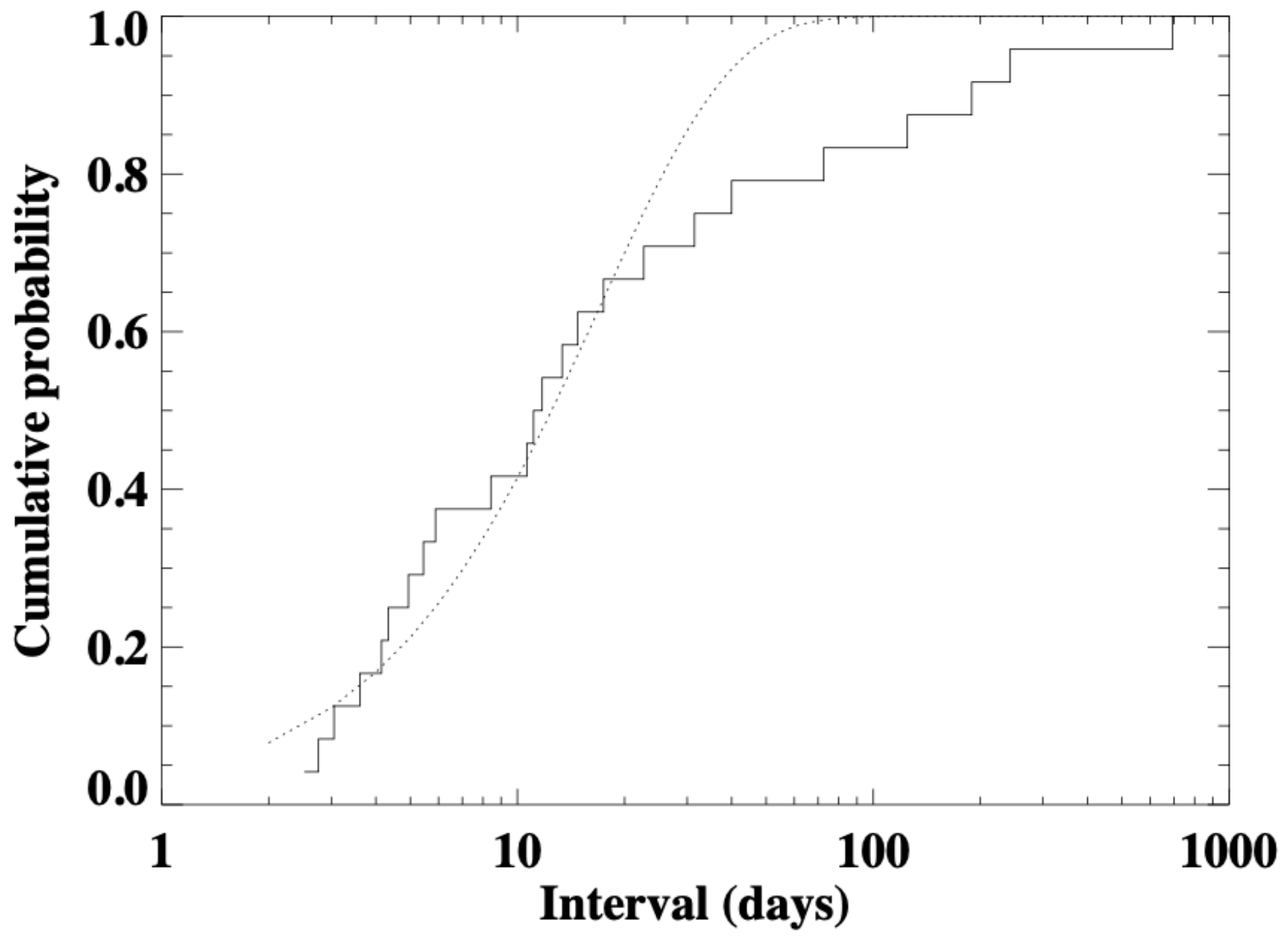

Figure 9 CDF of IPN20 and Weibull fit. The best fit (dotted line) to the intervals between bursts in our data greater than 2 days has scale and shape parameters of 17.07, and 1.169. The fit does not follow the features of the CDF well; only small sections of the CDF could be followed with a function of this form. In particular, the flattening from ~ 20-150 days, that reflects the transition from time scales within a window to those between windows, is not matched.

**Acknowledgments**

I thank Eric Linder and Kevin Hurley for valuable discussions, and the staff and scientists of the Energetic Cosmos Laboratory for support in this work. This paper has made use of data at the IPN website, http://ssl.berkeley.edu/ipn3, maintained by Kevin Hurley. This work was supported, in part, by the Energetic Cosmos Laboratory.

## 9. Appendix

Table A1

| Date | UT | IPNA | IPNB | KW+ |
|---|---|---|---|---|
| 05JUL14 | 09:32:48 | X | X | |
| 05JUL14 | 09:37:33 | X | X | |
| 05JUL14 | 09:41:05 | X | X | |
| 05JUL14 | 17:05:25 | X | X | |
| | | | | |
| 22FEB15 | 12:16:26 | X | X | |
| 22FEB15 | 12:31:11 | X | X | |
| 22FEB15 | 17:57:05 | X | X | |
| 22FEB15 | 19:44:16 | X | X | |
| 23FEB15 | 01:38:07 | X | X | |
| 23FEB15 | 05:24:54 | X | X | |
| 23FEB15 | 06:45:40 | X | X | |
| 23FEB15 | 12:27:52 | X | X | |
| 23FEB15 | 16:13:52 | X | X | |
| 24FEB15 | 06:48:47 | X | X | |
| 24FEB15 | 22:46:34 | X | X | |
| 25FEB15 | 10:51:03 | X | X | |
| 27FEB15 | 09:24:42 | X | X | |
| 28FEB15 | 10:56:48 | X | X | |
| 01MAR15 | 07:28:54 | X | X | |
| 05MAR15 | 18:59:19 | X | X | |
| 12APR15 | 11:24:24 | X | X | X |
| | | | | |
| 21DEC15 | 02:51:12 | X | X | |
| 01FEB16 | 09:37:15 | X | X | |
| | | | | |
| 14MAY16 | 08:21:54 | | X | |
| 16MAY16 | 20:49:46 | | X | |
| 18MAY16 | 08:37:07 | | X | |
| 18MAY16 | 09:09:23 | | X | X |
| 18MAY16 | 10:07:26 | | X | |
| 18MAY16 | 10:28:02 | | X | |
| 18MAY16 | 14:59:31 | | X | |
| 18MAY16 | 15:33:47 | | X | |
| 18MAY16 | 17:00:31 | | X | X |
| 18MAY16 | 19:40:37 | | X | |
| 19MAY16 | 00:34:15 | | X | |
| 19MAY16 | 12:07:46 | | X | |
| 20MAY16 | 05:21:33 | | X | X |
| 20MAY16 | 16:21:43 | | X | |
| 20MAY16 | 21:42:29 | | X | |
| 21MAY16 | 03:23:36 | | X | |
| 21MAY16 | 11:42:41 | | X | X |
| 21MAY16 | 20:01:47 | | X | |
| 21MAY16 | 20:23:42 | | X | |
| 21MAY16 | 20:43:09 | | X | |
| 21MAY16 | 21:13:16 | | X | |
| 22MAY16 | 00:12:59 | | X | |
| 02JUN16 | 12:19:30 | | X | |
| 17JUN16 | 06:27:57 | | X | X |
| 18JUN16 | 20:27:25 | | X | |
| 20JUN16 | 15:16:37 | | X | X |
| 21JUN16 | 11:07:38 | | X | |
| 22JUN16 | 13:45:23 | | X | |
| 23JUN16 | 12:58:17 | | X | |
| 23JUN16 | 15:16:09 | | X | |
| 23JUN16 | 16:49:57 | | X | |
| 23JUN16 | 17:21:09 | | X | X |
| 23JUN16 | 17:33:12 | | X | |
| 23JUN16 | 17:39:22 | | X | |
| 23JUN16 | 17:45:05 | | X | |
| 23JUN16 | 17:47:06 | | X | |
| 23JUN16 | 17:55:48 | | X | |
| 23JUN16 | 17:58:55 | | X | |
| 23JUN16 | 18:05:38 | | X | |
| 23JUN16 | 19:04:52 | | X | |
| 23JUN16 | 19:11:01 | | X | |
| 23JUN16 | 19:11:02 | | X | |
| 23JUN16 | 19:18:12 | | X | |
| 23JUN16 | 19:24:40 | | X | |
| 23JUN16 | 19:36:27 | | X | |
| 23JUN16 | 19:37:59 | | X | |
| 23JUN16 | 20:06:37 | | X | |
| 23JUN16 | 20:43:45 | | X | |
| 23JUN16 | 20:44:43 | | X | |
| 23JUN16 | 21:12:29 | | X | |
| 23JUN16 | 21:20:49 | | X | X |
| 23JUN16 | 21:23:36 | | X | |
| 23JUN16 | 22:16:37 | | X | |
| 23JUN16 | 22:18:15 | | X | |
| 23JUN16 | 22:23:30 | | X | |
| 23JUN16 | 22:39:51 | | X | |
| 23JUN16 | 22:41:58 | | X | |
| 25JUN16 | 08:04:52 | | X | |
| 25JUN16 | 20:32:19 | | X | |
| 26JUN16 | 05:15:20 | | X | |
| 26JUN16 | 06:03:15 | | X | |
| 26JUN16 | 09:40:15 | | X | X |
| 26JUN16 | 13:54:33 | | X | X |
| 26JUN16 | 14:22:11 | | X | |
| 26JUN16 | 16:07:08 | | X | |
| 26JUN16 | 17:22:55 | | X | |
| 26JUN16 | 17:50:03 | | X | |
| 26JUN16 | 18:22:27 | | X | X |
| 26JUN16 | 18:57:10 | | X | |
| 26JUN16 | 20:34:49 | | X | |
| 27JUN16 | 01:50:15 | | X | |
| 27JUN16 | 20:44:04 | | X | |
| 29JUN16 | 18:32:06 | | X | |
| 30JUN16 | 10:02:26 | | X | |
| 04JUL16 | 14:33:38 | | X | |
| 15JUL16 | 07:09:11 | | X | |
| 18JUL16 | 09:36:06 | | X | |
| 21JUL16 | 09:36:17 | | X | X |
| 05AUG16 | 04:51:17 | | X | |
| 25AUG16 | 07:26:53 | | X | |
| 26AUG16 | 14:05:52 | | X | |
| | | | | |
| 04OCT19 | 09:00:53 | X | X | |
| 29OCT19 | 13:39:16 | X | X | |
| 04NOV19 | 01:20:24 | X | X | |
| 04NOV19 | 01:54:37 | X | X | |
| 04NOV19 | 02:53:31 | X | X | |
| 04NOV19 | 04:26:55 | X | X | |
| 04NOV19 | 06:34:00 | X | X | |
| 04NOV19 | 09:17:53 | X | X | |
| 04NOV19 | 10:44:26 | X | X | X |
| 04NOV19 | 12:38:38 | X | X | |
| 04NOV19 | 14:47:29 | X | X | |
| 04NOV19 | 14:53:05 | X | X | |
| 04NOV19 | 15:36:47 | X | X | |
| 04NOV19 | 20:29:39 | X | X | |
| 04NOV19 | 23:48:01 | X | X | |
| 05NOV19 | 00:08:58 | X | X | |
| 05NOV19 | 01:36:25 | X | X | |
| 05NOV19 | 06:11:08 | X | X | X |
| 05NOV19 | 07:17:17 | X | X | |
| 15NOV19 | 20:48:41 | X | X | |
| | | | | |
| 10APR20 | 09:43:54 | X | | X |
| 22APR20 | 08:53:16 | X | | X |
| 27APR20 | 18:26:20 | X | | + |
| 27APR20 | 18:31:36 | X | | + |
| 27APR20 | 18:32:59 | X | | + |
| 27APR20 | 18:36:46 | X | | + |
| 27APR20 | 18:46:08 | X | | + |
| 27APR20 | 18:47:05 | X | | + |
| 27APR20 | 19:37:39 | X | | + |
| 27APR20 | 19:43:44 | X | | X |
| 27APR20 | 20:01:45 | X | | + |
| 27APR20 | 20:13:38 | X | | + |
| 27APR20 | 21:14:45 | X | | + |
| 27APR20 | 21:43:06 | X | | + |
| 27APR20 | 21:59:22 | X | | X |
| 27APR20 | 22:55:20 | X | | + |
| 27APR20 | 23:06:06 | X | | + |
| 27APR20 | 23:25:04 | X | | + |
| 27APR20 | 23:42:41 | X | | + |
| 27APR20 | 23:44:31 | X | | X |
| 28APR20 | 00:19:44 | X | | |
| 28APR20 | 00:39:39 | X | | |
| 28APR20 | 00:49:46 | X | | |
| 28APR20 | 01:04:03 | X | | |
| 28APR20 | 02:00:11 | X | | |
| 28APR20 | 03:47:52 | X | | |
| 28APR20 | 04:09:47 | X | | |
| 28APR20 | 05:56:30 | X | | |
| 28APR20 | 09:51:04 | X | | |
| *28APR20 | 14:34:24 | X | | X |
| 29APR20 | 20:47:27 | X | | |
| 03MAY20 | 23:25:13 | X | | |
| 10MAY20 | 06:12:02 | X | | X |
| 10MAY20 | 21:51:17 | X | | X |
| 19MAY20 | 18:32:30 | X | | |
| 20MAY20 | 14:10:49 | X | | |
| 20MAY20 | 21:47:07 | X | | |

Key: Bursts taken from IPN SGR list are given by UT time. All bursts in the table are in the IPN20 selection, but those in the IPNA or IPNB selections have an "X" in the appropriate column. In the KW+ column, an "X" or "+" indicates inclusion in the sample; an "X" showing an explicitly attributed K-W burst, a "+" denoting a burst consistent with GCN 27667 (Ridnaia, et al., 2020a), but not explicitly attributed to K-W in the IPN SGR list. (See text in Section 2.2 for explanation.) Blank rows indicate an apparent boundary between "episodes". The IPN SGB associated with FRB 20200428 is marked with an asterisk (Li, et al., 2021, CHIME/FRB Collaboration, 2020).